\documentclass[english]{article}
\usepackage[T1]{fontenc}
\usepackage[latin9]{inputenc}
\usepackage{amsmath}
\usepackage{graphicx}
\usepackage{amssymb}
\usepackage{esint}

\makeatletter

\providecommand{\tabularnewline}{\\}

\usepackage{amsxtra}\usepackage{ulem}

\usepackage{babel}

\makeatother

\usepackage{babel}

\begin{document}

\title{Relations between M-brane and D-brane quantum geometries}

\author{Josie Huddleston%
\thanks{Funded by an EPSRC studentship. With thanks to my supervisor, Douglas
Smith, and to everyone I've met within the Durham bubble. Particular
thanks to Sam and Eimear, who have both (in their turn) put up with
me during the worst moments, on my way through this stuff!%
}}

\date{28 June 2010}
\maketitle
\begin{abstract}
This paper investigates M-brane quantum geometry (and its representation
by equations involving the 3-bracket) by looking at the compactification
of an M-brane system to a D-brane system. Particularly of interest
is the system where coincident M2-branes end at an angle on an M5-brane,
and its reduction to D1- or F1-branes ending on a D3-brane. Since
equations for the quantum geometries of both of these systems are
known, the paper will attempt to directly relate them via a compactification.

\vfill{}
\pagebreak{}
\end{abstract}

\section{Introduction}

The M-branes of M-theory are not understood nearly so well as their
D-brane counterparts. Undoubtedly this has something to do with the
age of M-theory compared to that of string theory, but it is also
true to say that M-theory, as a higher dimensional theory and a generalisation,
has presented new challenges in areas that are now straightforward
for string theorists. One of these problems involves the quantisation
of N coincident D-branes, which in string theory has a low energy
limit based on a non-Abelian Yang-Mills Lie algebra. In M-theory,
the analogues are coincident M2-branes, which up until recently had
no simple description for their quantisation. Work by Bagger and Lambert
\cite{BagLam1}\cite{BagLam2}\cite{BagLam3}, and later Gustavsson
\cite{Gustavsson} shed light on this, using a 3-algebra in the same
sort of way as the multiple-F1 system used the Lie algebra, and hence
constructing an action (the Bagger-Lambert-Gustavsson action) for
coincident M2-branes.

However, the 3-algebra used by Bagger, Lambert and Gustavsson is itself
not well understood. When looking at two coincident M2s, the only
compatible 3-algebra found at the time had the simple 3-bracket $\left[T^{i},T^{j},T^{k}\right]=\varepsilon^{ijkl}T^{l}$
(although new examples are now coming to light \cite{new 3-algebrae}).
The difficulty in finding new solutions (as well as the requirement
for an invariant metric) is making the 3-bracket satisfy the Fundamental
Identity, the generalisation of the Jacobi Identity for a 3-algebra.

In this paper, we would like to shed some more light on the 3-bracket's
role in M-brane geometry. This task is aided in part by the work of
Chu and Smith, whose paper \cite{CS and Douglas} uses Basu-Harvey
fuzzy funnels in order to probe the $C$-field on the M5-brane with
multiple M2-branes to deduce properties of the M5 theory, similarly
to how multiple D1-branes can probe a $B$-field on the D3-brane.
In the latter case, this NS $B$-field translates via a Seiberg-Witten
map into the known non-commutative geometry on the D-brane represented
by $\left[X^{i},X^{j}\right]=\Theta^{ij}$ (for some $B$-field-related
constants $\Theta^{ij}$). For the 3-algebra, in contrast, the $C$-field
leads to an equation $\left[X^{i},X^{j},X^{k}\right]=\Theta^{ijk}$,
which represents some novel quantum geometry for the M5-brane.

To get more information about the 3-bracket (both its form and its
meaning within the theory), it would be useful to know how it relates
to the commutator - i.e. to reduce the M5 action to, say, a D3 action
and hence (by comparison with known D3 actions) relate the constants
$\Theta^{ijk}$ from the M-brane geometry to the constants $\Theta^{ij}$.
This would begin to give us some insight into how the quantum geometry
of the M5-brane works.

The results of two papers are useful here. The previously mentioned
paper \cite{CS and Douglas}, of course, provides the link between
the 3-bracket equation and the $C$-field. The second paper, by Pasti,
Sorokin and Tonin \cite{main compactif paper} provides an action
for an M5-brane with a $C$-field, and a reduction as far as a D4-brane
for a simplified case.

The full reduction using a $C$-field with all components switched
on will be beyond the scope of this paper, but instead we will focus
on the two-component case considered in \cite{CS and Douglas} where
the coincident M2-branes end on an M5-brane at an angle $\alpha$.
This will involve following the process of \cite{main compactif paper}
but without setting $A_{\mu5}$ to zero. A key question is whether
the results of this compactification will tally with known results
about coincident strings ending on D3-branes at an angle, and in the
final part of the paper a comparison with the non-commutative geometry
known for D-branes will be attempted to assess this, using results
from \cite{problem paper}.

\section{\label{sec:The-compactification}The compactification}

We consider an M2-brane with coordinates 0 and 2 in common with the
M5-brane that it connects with:\medskip{}

\begin{tabular}{|c|c|c|c|c|c|c|c|}
\hline 
M2:  & 0  &  & 2  &  &  &  & 6\tabularnewline
\hline
\hline 
M5:  & 0  & 1  & 2  & 3  & 4  & 5  & \tabularnewline
\hline
\end{tabular}

\subsection{Dualising the M5 action}

We begin with the known M5 action, from \cite{first M5 paper}.

\begin{multline}
S_{\textrm{M5}}=\int\textrm{d}^{6}x\,\left(\sqrt{-\det\left(g_{mn}+i\bar{H}_{mn}\right)}+\frac{\sqrt{-g}}{4}\partial_{m}a(\star H)^{mnl}H_{nlp}\partial^{p}a\right)+\\
\int\left(C^{(6)}+\frac{1}{2}F^{(3)}\wedge C^{(3)}\right)\label{eq:M5M2}\end{multline}
where $H=F-C$ ($F=\textrm{d}A$ the worldvolume field strength),
$\star H$ is the Hodge star of H, and $\bar{H}_{mn}=\frac{1}{\sqrt{(\partial a)^{2}}}(\star H)_{mnl}\partial^{l}a$.
The field $a$ is an auxiliary field introduced to ensure d=6 covariance
of the action \cite{first M5 paper}, since the self-dual field-strength
$H_{(3)}$ on its own would prevent us from writing a kinetic term
(since $H\wedge\star H=H\wedge H=0$). In order for $S_{\textrm{M5}}$
to reduce to a standard D4 action, it is also necessary to dualise
it. \cite{main compactif paper} contains a reduction/dualisation
of the M5 action, but they make one assumption that we don't want
to make here, namely that $A_{\mu r}=0$ (where $r$ is the direction
of reduction). We retain the assumption of a gauge choice such that
the auxiliary field $a$ satisfies $\partial_{\mu}a=\delta_{\mu}^{r}$,
and the assumption of a metric whose determinant is unchanged after
reduction, specifically in directions 2 and 5.

The removal of this assumption causes terms involving $F_{mn5}^{(3)}=\left(\textrm{d}A\right)_{mn5}$
to become non-zero. This means that any direction of reduction will
result in non-zero fields $F^{(2)}$ as well as the $F^{(3)}$ fields
found in \cite{main compactif paper}. This in turn makes dualisation
of the fields rather more complicated - before, the $F$-fields could
be dualised by replacement with their Hodge star (give or take a constant
factor) but this is only true for $A_{\mu r}=0$ and compactification
in the $r$ direction.

To do the dualisation from scratch, we begin by expanding some of
the $H$s, and splitting $F$ and $C$ into their 2- and 3-form parts:

\begin{multline*}
S\left[F^{(2)},F^{(3)}\right]=\int\textrm{d}^{5}x\,\sqrt{\det\left(g_{\alpha\beta}+i(\star H)_{\alpha\beta}\right)}+\epsilon^{\alpha\beta\gamma\delta\epsilon}C_{\alpha\beta\gamma\delta\epsilon}^{(5)}+\qquad\qquad\qquad\\
\epsilon^{\alpha\beta\gamma\delta\epsilon}\left(-\tfrac{1}{24}\left(F^{(2)}F^{(3)}+C^{(2)}C^{(3)}\right)_{\alpha\beta\gamma\delta\epsilon}+\tfrac{1}{12}\left(F^{(2)}C^{(3)}+C^{(2)}F^{(3)}\right)_{\alpha\beta\gamma\delta\epsilon}\right)\end{multline*}
The next step is to add two different Lagrange multiplier terms to
the action in order to impose the $F=\textrm{d}A$ constraint in both
the 2- and 3-form cases \cite{Lagrangemulthelp}:

\begin{multline*}
\hat{S}\left[F^{(2)},F^{(3)},\star\hat{F}^{(2)},\star\hat{F}^{(3)}\right]=S\left[F^{(2)},F^{(3)}\right]+\qquad\\
\int\epsilon^{\alpha\beta\gamma\delta\epsilon}\left(\tfrac{1}{2}i\left(\star\hat{F}\right)_{\alpha\beta\gamma}^{(3)}\left(F_{\delta\epsilon}^{(2)}-2\partial_{\delta}A_{\epsilon}\right)+\tfrac{1}{6}i\left(\star\hat{F}\right)_{\alpha\beta}^{(2)}\left(F_{\gamma\delta\epsilon}^{(3)}-2\partial_{\gamma}A_{\delta\epsilon}\right)\right)\end{multline*}
The two Lagrange multipliers are $\tfrac{1}{2}i\left(\star\hat{F}\right)_{\alpha\beta\gamma}^{(3)}$
and $\tfrac{1}{6}i\left(\star\hat{F}\right)_{\alpha\beta}^{(2)}$,
with their constants chosen for convenience. Note that despite the
factors of $i$, these terms will turn out to be real.

From here, taking equations of motion in each of $F^{(3)}$ and $F^{(2)}$
in turn, results in two independent equations that determine the new
$F$s in terms of the old:\[
\hat{F}^{(2)}=\frac{i}{2}\tilde{C}^{(2)}-\frac{i}{4}\tilde{F}^{(2)}\qquad\qquad\hat{F}^{(3)}=\frac{i}{6}\tilde{C}^{(3)}-\frac{i}{12}\tilde{F}^{(3)}\]
where tildes indicate Hodge stars. Finally, in order to keep the Born-Infeld
part of the new action in the same form as the old one, we define
the new $\hat{C}$-fields by saying that $\hat{H}^{(2)}=k\left(i\star H^{(3)}\right)$
(where $\hat{H}^{(2)}=\hat{F}^{(2)}-\hat{C}^{(2)}$ analogous to $H=F-C$)
. This results in:\[
\hat{C}^{(2)}=\frac{i}{12}\star C^{(3)}\qquad\qquad\hat{C}^{(3)}=i\star C^{(2)}\]

Substituting into \eqref{eq:M5M2}, the result is the dualised-reduced-M5
action:

\begin{multline}
\hat{S}=\int\textrm{d}^{5}x\,\sqrt{\det\left(g_{\alpha\beta}+12i\hat{H}_{\alpha\beta}\right)}+\epsilon^{\alpha\beta\gamma\delta\epsilon}\left(\hat{C}_{\alpha\beta\gamma\delta\epsilon}^{(5)}+2\hat{F}_{\alpha\beta\gamma}^{(3)}\hat{F}_{\delta\epsilon}^{(2)}\right)\end{multline}
 where we have absorbed the $\hat{C}^{(2)}\hat{C}^{(3)}$ term along
with $C^{(5)}$ into the new $\hat{C}^{(5)}$. Note that this is a
D4 action, but we haven't chosen which numbers the indices run over
yet. We do so in the next section.

\subsection{\label{sub:secwithCfields}The D2-on-D4 and F1-on-D4 actions}

In order to simplify things a bit for the purposes of more reduction,
we'll start with the above (dualised-reduced-)M5 action with a specific
$C$-field which has two non-zero parts: one including the direction
of the M2-brane and one not. Specifically, we'll look at the choice
found in \cite{CS and Douglas} which represents an M2-brane at an
angle of $\alpha$ to the M5-brane, i.e. $\tilde{C}_{012}^{(3)}=\tfrac{1}{4}\sin\alpha$,
$\tilde{C}_{345}^{(3)}=-\tfrac{1}{4}\tan\alpha$ (with all other $\tilde{C}^{(3)}$
and also $\tilde{C}^{(5)}$ zero). We then have two choices of how
to compactify down to a D4-brane in IIA - by compactifying a direction
perpendicular to the M2-brane (say 5), or a dimension parallel to
the M2-brane (say 2). These produce a D2 on a D4, and an F1 on a D4
respectively, with actions as shown here.

\begin{multline}
S_{(\textrm{D4,D2})}=\int\textrm{d}^{5}x\,\sqrt{\det\left(g_{\alpha\beta}+\tilde{F}_{\alpha\beta}^{(2)}\right)-\sin\alpha}+\frac{i}{24}\epsilon^{\alpha\beta\gamma\delta\epsilon}\tilde{F}_{\alpha\beta\gamma}^{(3)}\tilde{F}_{\delta\epsilon}^{(2)}\\
-\frac{i}{12}\tilde{F}_{34}^{(2)}\tan\alpha-\frac{i}{12}\tilde{F}_{012}^{(3)}\sin\alpha+\frac{i}{6}\sin\alpha\tan\alpha\end{multline}

\begin{multline}
S_{(\textrm{D4,F1})}=\int\textrm{d}^{5}x\,\sqrt{\det\left(g_{\alpha\beta}+\tilde{F}_{\alpha\beta}^{(2)}\right)-\tan\alpha}+\frac{i}{24}\epsilon^{\alpha\beta\gamma\delta\epsilon}\tilde{F}_{\alpha\beta\gamma}^{(3)}\tilde{F}_{\delta\epsilon}^{(2)}\\
-\frac{i}{12}\tilde{F}_{01}^{(2)}\sin\alpha-\frac{i}{12}\tilde{F}_{345}^{(3)}\tan\alpha+\frac{i}{6}\sin\alpha\tan\alpha\end{multline}
Note also that the original two $C$-field components come through
this unchanged; to get a result for a more general two-component $C$-field
is simply a matter of replacing $\tan\alpha$ and $\sin\alpha$ above
respectively with $\tilde{C}_{012}^{(3)}$ and $\tilde{C}_{34}^{(2)}$
(D4-D2 case) or $\tilde{C}_{345}^{(3)}$ and $\tilde{C}_{01}^{(2)}$
(D4-F1 case).

\subsection{The D1-on-D3 and F1-on-D3 actions}

To turn the D4 actions into their related D3 actions we must T-dualise,
in directions 2 and 5 respectively \cite{johnson}. Since all the
WZ terms have a component of these directions (recall that $\sin\alpha$
and $\tan\alpha$ are $C$-fields in the directions perpendicular
to the $F$s) this simply has the effect of removing all 2's and 5's
and reducing the dimensions by 1. (Note that the Born-Infeld part
of each action remains the same.)

\begin{multline}
S_{(\textrm{D3,D1})}=\int\textrm{d}^{4}x\,\sqrt{\det\left(g_{\alpha\beta}+\tilde{F}_{\alpha\beta}^{(2)}\right)-\sin\alpha}+\frac{i}{24}\epsilon^{\alpha\beta\gamma\delta}\left(\left(\tilde{F}_{\alpha\beta\gamma}^{(3)}\tilde{F}_{\delta}^{(1)}\right)+\tilde{F}_{\alpha\beta}^{(2)}\tilde{F}_{\gamma\delta}^{(2)}\right)\\
-\frac{i}{12}\tilde{F}_{34}^{(2)}\tan\alpha-\frac{i}{12}\tilde{F}_{01}^{(2)}\sin\alpha+\frac{i}{6}\sin\alpha\tan\alpha\label{eq:D3D1}\end{multline}

\begin{multline}
S_{(\textrm{D3,F1})}=\int\textrm{d}^{4}x\,\sqrt{\det\left(g_{\alpha\beta}+\tilde{F}_{\alpha\beta}^{(2)}\right)-\tan\alpha}+\frac{i}{24}\epsilon^{\alpha\beta\gamma\delta}\left(\left(\tilde{F}_{\alpha\beta\gamma}^{(3)}\tilde{F}_{\delta}^{(1)}\right)+\tilde{F}_{\alpha\beta}^{(2)}\tilde{F}_{\gamma\delta}^{(2)}\right)\\
-\frac{i}{12}\tilde{F}_{01}^{(2)}\sin\alpha-\frac{i}{12}\tilde{F}_{34}^{(2)}\tan\alpha+\frac{i}{6}\sin\alpha\tan\alpha\label{eq:D3F1}\end{multline}
Again, the general two-component $C$-field result follows by directly
substituting for sin and tan alpha, crossing out the 2s and 5s from
the general $C$-fields given at the end of the last section. From
this, and the standard D3 action found in \cite{johnson}, we can
see that in each case one of the two parts of the $C$-field has become
a RR $C^{(2)}$-field and one part has become a NS $B^{(2)}$-field.
Specifically, we get $C_{01}^{(2)}=\tan\alpha$, $B_{34}=C_{34}^{(2)}=\sin\alpha$
for the D3-D1 case, and $C_{34}^{(2)}=\sin\alpha$, $B_{01}=C_{01}^{(2)}=\tan\alpha$
for the D3-F1 case.

\subsection{\label{sub:Geometry}A word about geometry}

At this point, it is worth taking a moment to consider how all this
relates back to the geometry of the D- and M-brane systems. We are
looking particularly at set-ups where a lower dimensional brane ends
on a higher dimensional brane at an angle, and have established that
it's possible to reduce and dualise in such a way as to turn the M5-M2
system into either a D3-D1 system or a D3-F1 system. In terms of the
D-brane geometry, we know that adding a constant NS $B$-field to
the D-brane worldvolume causes the worldvolume coordinates $X^{i}$
to obey a commutation relation\begin{equation}
\Theta^{ij}=i\left[X^{i},X^{j}\right]\label{eq:commutator}\end{equation}
 with components of $\Theta$ dependent on the $B$-field. For small
$B$, this can be written as \begin{equation}
(2\pi\alpha^{\prime})^{2}\mathcal{F}^{ij}=i\left[X^{i},X^{j}\right]\label{eq:commutatorappr}\end{equation}
 where $\mathcal{F}=F+B$ and $F$ is the worldvolume field strength
on the brane \cite{CS and Douglas}. Note that since the geometry
depends only on total field $\mathcal{F}$, it is indifferent to which
parts of the field come from the field strength $F$ and which from
the $B$-field.

In the M-brane case we expect an analogous situation with the quantum
geometry represented by the 3-bracket - that is, upon adding a $C^{(3)}$-field
it takes the form\begin{equation}
\Theta^{ijk}\sim i\left[X^{i},X^{j},X^{k}\right]\label{eq:3bracket}\end{equation}
 and if $C$ is sufficiently small we can approximate this by \begin{equation}
\mathcal{F}^{ijk}\sim i\left[X^{i},X^{j},X^{k}\right]\label{eq:3bracketappr}\end{equation}
 with $\mathcal{F}=F+C^{(3)}$ \cite{CS and Douglas}. In the reduction,
as we've seen, this $C^{(3)}$-field splits into two pieces, with
one part becoming a RR $C^{(2)}$-field and one part becoming a NS
$B^{(2)}$-field. Now in equation \eqref{eq:commutatorappr}, we have
$\mathcal{F}=F_{0}+B+C^{(2)}$ (with $F_{0}$ the original worldvolume
field strength). Hence it is clear that in theory the same D-brane
geometry could be produced by various sets of fields, so long as they
come together to the same $\mathcal{F}$.

From a purely geometric standpoint, we can see that reducing and dualising
an angled-M2-on-M5 system should produce an angled-D1/F1-on-D3 system,
with the same angle $\alpha$ involved throughout. Thus, we might
expect that the geometry as specified by \eqref{eq:3bracket} should
reduce to the geometry as specified by \eqref{eq:commutator}. The
question is, can we explicitly check this for the particular case
we're looking at? From \cite{CS and Douglas}, we know that\begin{equation}
\left[X^{\mu},X^{\nu},X^{\lambda}\right]=\begin{cases}
\frac{i\varepsilon^{\mu\nu\lambda}\sin\alpha}{K\cos^{4}\alpha} & \mu,\nu,\lambda\in\{0,1,2\}\\
-\frac{i\varepsilon^{\mu\nu\lambda}\tan\alpha}{K\sec^{4}\alpha} & \mu,\nu,\lambda\in\{3,4,5\}\\
0 & \textrm{otherwise}\end{cases}\label{eq:3-bracketdetail}\end{equation}
 (where $K$ is a dimensional proportionality constant) is the relevant
equation for the angled-M2-on-M5 (with $C$-fields as specified in
\ref{sub:secwithCfields}), while\begin{equation}
\left[X^{\mu},X^{\nu}\right]=\begin{cases}
\frac{i\varepsilon^{\mu\nu}(2\pi\alpha^{\prime})\tan\alpha}{1+\tan^{2}\alpha} & \mu,\nu\in\{3,4\}\\
0 & \textrm{otherwise}\end{cases}\label{eq:commdetail}\end{equation}
 is the equation for an angled-D1-on-D3, arrived at by including a
$B$-field $B_{34}^{(2)}=\frac{\tan\alpha}{(2\pi\alpha^{\prime})}$
\cite{tilting}. Thus we make the {}``Geometry Reduction conjecture''
that the system with geometry specified by \eqref{eq:3-bracketdetail}
should reduce to one with geometry specified by \eqref{eq:commdetail}.

\section{Checking the Geometry Reduction conjecture}

So far, we have found the explicit reduction of $C^{(3)}$ yields
a $B$ and $C^{(2)}$ combination. For example, in the D3-D1 case
\eqref{eq:D3D1}, we get $C_{01}^{(2)}\sim\tan\alpha$ plus $B_{34}=C_{34}^{(2)}\sim\sin\alpha$.
This clearly is not the same as the standard example of an angled
D3-D1 system found in the literature, which has only a $B$-field
$B_{34}^{(2)}\sim\tan\alpha$. However, as we showed in \ref{sub:Geometry},
it is entirely possible that two such systems could indeed have the
same non-commutative geometry, and indeed there is good reason to
suggest that they should. To check this, though, it is necessary to
construct the total field strength from the various fields obtained
in the reduction.

In \cite{problem paper}, Cornalba et al. give a method for constructing
a field strength $\mathcal{F}$ from a background $U(1)$ gauge field
strength $F_{0}$ with infinitesimal $\delta B$- and $\delta C$-fields
added to it using\begin{equation}
\delta\mathcal{F}=\frac{1}{2}\left(1+F_{0}\frac{1}{g}\right)\Omega\left(1-\frac{1}{g}F_{0}\right)\label{eq:addafield}\end{equation}
where 

\[
\Omega=\frac{1}{1+F_{0}\tfrac{1}{g}}\left(\delta B-F_{0}\frac{1}{g}\delta B\frac{1}{g}F_{0}\right)\frac{1}{1-\tfrac{1}{g}F_{0}}\]
for the $B$-field, and

\[
\Omega_{mn}\gamma^{mn}=\left.e^{-\omega\cdot\gamma}\delta C\gamma^{4}\ldots\gamma^{9}-\overline{\delta C}\gamma^{4}\ldots\gamma^{9}e^{\omega\cdot\gamma}\right|_{\textrm{2-form}}\]
for the $C$-field, with $\omega\cdot\gamma=\omega_{mn}\gamma^{m}\gamma^{n}$
and $\omega$ a specific 2-form that depends on the initial $F_{0}$.

It would be useful to know whether this method of deriving $\mathcal{F}$
is still applicable outside the infinitesimal domain, i.e. for general
$B$ and $C$. Certainly there is an obvious way to proceed, by calculating
$\delta\mathcal{F}$ in terms of $\delta B$ or $\delta C$ and then
solving the resulting differential equation(s) to find the new $\mathcal{F}$
in terms of fields $F_{0}$, $B$ and $C$. In a footnote of \cite{problem paper}
this is done for a finite $B$-field with $C$ and $F_{0}$ set to
zero, and the answer is $\mathcal{F}=g\tanh\left(\frac{1}{2}\frac{1}{g}B\right)$.%
\begin{figure}[bh]
\begin{centering}
\includegraphics[bb=0bp 115bp 240bp 235bp,clip]{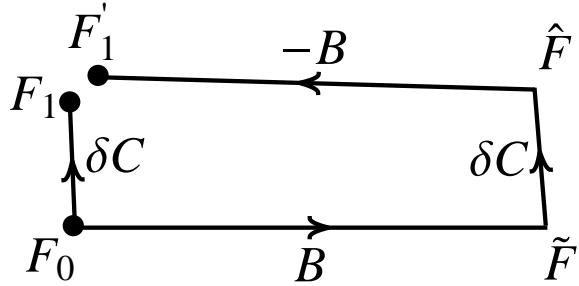}
\par\end{centering}

\caption{\label{fig:finiteinf}Finite $B$-field, infinitesimal $\delta C$-field}

\end{figure}

In order for the method to make sense for finite order though, it
is necessary for it to be path independent. That is to say, it shouldn't
matter in what order the finite $B$ and $C$ are added (or even if
they are added piece by piece) - the resulting $\mathcal{F}$ should
be the same. A first step towards this is to look at putting together
a finite $B$-field with an infinitesimal $\delta C$-field, such
as shown in Figure \ref{fig:finiteinf}. There are two paths on this
diagram - one that takes the simple route from $F_{0}$ to $F_{1}$
by the addition of an infinitesimal $\delta C$, and one that takes
the long way round to $F_{1}^{\prime}$ by adding a finite $B$, then
the $\delta C$, then $-B$. Note that the process of putting a new
field into $\mathcal{F}$ is not really as simple as addition, so
it is a nontrivial problem to ask whether $F_{1}$ and $F_{1}^{\prime}$
are the same, or to what order in $\delta C$ equality holds. (If
equality holds for sufficiently high order, it should be possible
to {}``sum up'' (i.e. integrate) a block of such paths to prove
path-independence for finite $B$ and $C$.)

Checking this for general $B$ and $\delta C$ turns out to be rather
difficult. When all components of the two fields are switched on the
result is a matrix of interlinked differential equations with no easy
solutions. However, with the results of Section \ref{sec:The-compactification}
in mind it would be instructive to look at the case where $\delta C=\delta C_{01}$,
$B=B_{34}$. With only one component in each field, equation \eqref{eq:addafield}
can be solved repeatedly to find the new field. First to find $F_{1}$:\begin{multline}
\qquad\qquad\qquad\qquad e^{\pm\omega\cdot\gamma}=\frac{1}{\sqrt{1+F^{2}}}\pm\frac{F}{\sqrt{1+F^{2}}}\gamma^{0}\gamma^{1}\\
\delta F=\tfrac{1}{2}\delta C\,\sqrt{F^{2}+1}\qquad\Longrightarrow\qquad F_{1}=F_{0}+\tfrac{1}{2}\delta C\,\sqrt{F_{0}^{2}+1}+O\left((\delta C)^{2}\right)\label{eq:F1}\end{multline}
Then $\tilde{F}$:\[
\delta F=\tfrac{1}{2}\delta B\,\left(1+F^{2}\right)\qquad\Longrightarrow\qquad\tfrac{1}{2}B=\intop_{F_{0}}^{\tilde{F}}\frac{\textrm{d}F}{F^{2}+1}\qquad\Longrightarrow\qquad\tilde{F}=\frac{\tan\tfrac{1}{2}B+F_{0}}{1-F_{0}\tan\tfrac{1}{2}B}\]
Then $\hat{F}$:\[
\hat{F}=\tilde{F}+\tfrac{1}{2}\delta C\,\sqrt{\tilde{F}^{2}+1}+O\left((\delta C)^{2}\right)\]
And finally $F_{1}^{\prime}$:\[
-\tfrac{1}{2}B=\intop_{\hat{F}}^{F_{1}^{\prime}}\frac{\textrm{d}F}{F^{2}+1}\qquad\Longrightarrow\qquad F_{1}^{\prime}=\frac{-\tan\tfrac{1}{2}B+\hat{F}}{1+\hat{F}\tan\tfrac{1}{2}B}\]
After substituting and simplifying:\[
F_{1}^{\prime}=F_{0}+\tfrac{1}{2}\delta C\,\sqrt{1+F_{0}^{2}}\left(\cos\tfrac{1}{2}B-F_{0}\sin\tfrac{1}{2}B\right)+O\left((\delta C)^{2}\right)\]
This agrees to constant term, but no further, with the $F_{1}$ result
in \eqref{eq:F1}. It also agrees to first order and beyond in $\delta C$
if $B$ and $F_{0}$ are infinitesimal, in line with the results of
\cite{problem paper} which says that $F_{1}$ and $F_{1}^{\prime}$
should be equal in this case.

This unfortunately suggests that the $\mathcal{F}$-construction method
will not work for general $B$ and $C$. Hence it cannot reliably
be used to check the Geometry Reduction conjecture made at the end
of Section \ref{sec:The-compactification}.

\section{Conclusions}

We have seen that it is possible to generalise the results of \cite{main compactif paper}
to the case where $A_{\mu r}\neq0$. The continued assumption of gauge
choice such that $\partial_{\mu}a=\delta_{\mu}^{r}$ seems reasonable,
while it might be interesting (if rather trickier) in future work
to look at breaking the final assumption: i.e. to look at some specific
cases where the determinant of the metric contributes extra terms
to the action when reduced in directions 2 and/or 5.

In this paper, the generalisation has led to some new D4 and D3 actions
representing various possible compactifications of an M5 with background
$C$-field. In particular, by comparison with well-known actions for
the D3-brane, we have seen that under these conditions a $C^{(3)}$-field
with two orthogonal components reduces to two 2-forms, $B$ and $C^{(2)}$,
with directly comparable components to those of the original field.
These have suggested that the non-commutative geometry of the D3-brane
could be arrived at using a combination of $B$- and $C$-fields,
as well as purely from a $B$-field as we are accustomed to. If proved
fully, this illumination could work both ways: the method could be
useful in Bagger-Lambert theory, if the poorly-understood 3-bracket
used to describe the quantum geometry of the M5-brane could be in
some way directly compared to the simpler commutator used for D3-branes.
Alternatively, finding that this particular $B$ and $C$ combination
\textit{does} in fact result in a different D-brane non-commutative
geometry (i.e. that the conjecture is false) would also be an interesting
result, though perhaps it is less likely given what we know from the
geometrical picture.

In the last section of the paper, we saw that the method for constructing
a single field $\mathcal{F}$ out of background fields (found in \cite{problem paper})
was path-dependent, and hence badly-defined, for non-infinitesimal
combinations of $B$ and $C$ fields. Any future work modifying this
method for finite fields should be able to use the results of this
paper to check the Geometry Reduction conjecture, i.e. to see for
sure whether the reduction of an M-brane quantum geometry is (at least
in this specific case) equivalent to the known D-brane non-commutative
geometry. If this is indeed the case, the next step would be to try
and get some idea of what happens to the 3-bracket during the reduction
process, as this might lead to a method to convert any known $\Theta^{ijk}(\alpha)$
(i.e. any specific instance of \eqref{eq:3bracket}), into the $\Theta^{ij}(\alpha)$
that specifies the non-commutative geometry of the reduced system.

\end{document}